\documentclass[aps,prd,10pt,nofootinbib,preprintnumbers,twocolumn,superscriptaddress]{revtex4-1}
\usepackage{amsmath,amssymb,amsfonts,dsfont,mathrsfs,amsthm}
\usepackage{graphicx}
\usepackage{hyperref}
\hypersetup{linktocpage,colorlinks=true,urlcolor=blue,linkcolor=blue,citecolor=blue}
\usepackage{siunitx}
\usepackage[paperwidth=612pt,paperheight=792pt,top=28pt,right=34pt,bottom=35pt,left=34pt]{geometry}

\usepackage{bigints}
\usepackage[english]{babel}
\usepackage{float}
\usepackage{enumitem}
\usepackage{array}

\newcommand{\diff}[1]{\text{d}#1}

\newcommand{\Lag}{\mathscr{L}}

\newcommand{\dv}{\diff{}^4x\sqrt{-g}}

\newcommand*{\arcsinh}{\operatorname{arcsinh}}

\begin{document}

\title{Diffusion in unimodular gravity: Analytical solutions, late-time acceleration, and cosmological constraints}

\author{Crist\'obal Corral}
\email{crcorral@unap.cl}
\affiliation{Instituto de Ciencias Exactas y Naturales, Facultad de Ciencias, Universidad Arturo Prat, Avenida Arturo Prat Chac\'on 2120, 1110939, Iquique, Chile}
\affiliation{Departamento de F\'isica, Universidad de Santiago de Chile, Avenida Ecuador 3493, Estaci\'on Central, 9170124, Santiago, Chile}

\author{Norman Cruz}
\email{norman.cruz@usach.cl}
\affiliation{Departamento de F\'isica, Universidad de Santiago de Chile, Avenida Ecuador 3493, Estaci\'on Central, 9170124, Santiago, Chile}

\author{Esteban Gonz\'alez}
\email{esteban.gonzalezb@usach.cl}
\affiliation{Departamento de F\'isica, Universidad de Santiago de Chile, Avenida Ecuador 3493, Estaci\'on Central, 9170124, Santiago, Chile}

\begin{abstract}
    Unimodular gravity is an appealing approach to address the cosmological constant problem. In this scenario, the vacuum energy density of quantum fields does not gravitate and the cosmological constant appears merely as an integration constant. Recently, it has been shown that energy diffusion that may arise in quantum gravity and in theories with spontaneous collapse is compatible with this framework by virtue of its restricted diffeomorphism invariance. New studies suggest that this phenomenon could lead to higher-order equations in the context of homogeneous and isotropic Universe, affecting the well-posedness of their Cauchy initial-value problem. In this work, we show that this issue can be circumvented by assuming an equation of state that relates the energy density to the function that characterizes diffusion. As an application, we solve the field equations analytically for an isotropic and homogeneous Universes in a barotropic model and in the mass-proportional continuous spontaneous localization (CSL) scenario, assuming that only dark matter develops energy diffusion. Different solutions possessing phase transition from decelerated to accelerated expansion are found. We use cosmological data of type Ia Supernovae and observational Hubble data to constrain the free parameters of both models. It is found that very small but nontrivial energy nonconservation is compatible with the barotropic model. However, for the CSL model, we find that the best-fit values are not compatible with previous laboratory experiments. We comment on this fact and propose future directions to explore energy diffusion in cosmology.
\end{abstract}

\maketitle

\section{Introduction}

Independent observations of type Ia Supernovae (SNe Ia) suggest that the Universe is passing through a phase of accelerated expansion~\cite{Riess1998,Perlmutter1999}; an event that it is likely to have begun at a redshift of $z=0.64$~\cite{Hinshaw_Nine_2013,Ade:2015xua,Moresco:2016mzx}. The simplest explanation for this phenomenon is to add the cosmological constant to the Einstein's field equations. When matter fields are included, however, quantum fluctuations generate an effective value that it tightly constrained by observational data. This fact pushes down the bare cosmological constant by several orders of magnitude in order to cancel the huge contributions coming from the matter sector. The lack of explanation for this unnatural choice of the parameters is known as the cosmological constant problem (for a review see~\cite{Weinberg:1988cp}).

On the other hand, the discrepancy between two independent measurements of the Hubble constant that disagree at four sigma level can be regarded as a current cosmological puzzle as well~\cite{Addison:2017fdm,Macaulay:2018fxi,Wong:2019kwg}. The first one involves the measurement of the receding velocity of SNe Ia and it represents a model-independent estimation~\cite{Riess:2016jrr,Riess:2018byc,Riess:2019cxk}. The second one, is an extrapolation of the data coming from the cosmic microwave background (CMB) and it assumes the Lambda cold dark matter ($\Lambda$CDM) paradigm~\cite{Aghanim:2018eyx}. There are three possible resolutions to this tension: (i)~a modification of the statistical setup to refine the measurements~\cite{Bennett:2014tka,Cardona:2016ems,Feeney:2017sgx,Zhang:2017aqn}, (ii)~possible systematic errors that have not been fully accounted for the analysis, and (iii)~a modification of the laws of gravitation.

Unimodular gravity (UG) is an interesting approach to deal with these two puzzles at once. It is based on a restricted diffeomorphism invariance of the Einstein--Hilbert action that preserves the volume element~\cite{Einstein:1919gv,Anderson:1971pn,vanderBij:1981ym,Buchmuller:1988wx,Unruh:1988in,Henneaux:1989zc,Ng:1990xz,Finkelstein:2000pg,Ellis:2010uc}. In fact, several of their properties and implications at the quantum level have been studied~\cite{Alvarez:2005iy,Alvarez:2006uu,Jain:2012gc,Barvinsky:2017pmm,Alvarez:2008zw,Smolin:2009ti,Fiol:2008vk,Barcelo:2014mua,Padilla:2014yea,Saltas:2014cta,Alvarez:2015pla,Alvarez:2015sba,Bufalo:2015wda,Eichhorn:2015bna,Benedetti:2015zsw,Josset:2016vrq,Ardon:2017atk,Percacci:2017fsy}.\footnote{There exist, however, a covariant formulation that introduces a dynamical exact $4$-form, whose value is fixed on-shell by a Lagrange multiplier~\cite{Henneaux:1989zc}. Later, it was shown that one can avoid introducing the latter and still have a generally covariant theory~\cite{Jirousek:2018ago,Hammer:2020dqp}.} Their analysis through constrained Hamiltonian dynamics shows that, even though it has less symmetries than general relativity (GR), it propagates the same number of degrees of freedom by virtue of an additional constraint~\cite{Henneaux:1989zc}. 

The Noether theorem associated to this symmetry implies a modified conservation law for matter fields. On shell, Bianchi identities give rise to the Einstein's field equation with the cosmological constant arising as an integration constant. Moreover, the vacuum energy density of quantum fields can be removed from the field equations by rescaling an additional component of the energy-momentum tensor that appears from the restricted invariance~\cite{Smolin:2009ti}. In consequence, there is no cosmological constant problem whatsoever. It is worth mentioning that this reduced symmetry has been explored in modified gravity theories as well, e.g. in $f(R)$ and $f(T)$ gravities~\cite{Nojiri:2015sfd,Bamba:2016wjm,Nassur:2016yhc,Rajabi:2017alf}, $F(\mathcal{G})$ with $\mathcal{G}$ being the Gauss--Bonnet invariant~\cite{Houndjo:2017jsj}, unimodular Einstein--Cartan theory~\cite{Alvarez:2015oda,Bonder:2018mfz}, supergravity~\cite{Anero:2019ldx,Anero:2020tnl}, to mention a few.    

On the other hand, UG provides a suitable setup to reconcile gravitation with the energy-momentum nonconservation that arises from quantum collapse or quantum gravity discreteness~\cite{Josset:2016vrq,Perez:2017krv,Perez:2018wlo}. As mentioned in Ref.~\cite{Josset:2016vrq}, this phenomenon appears when the density matrix is modeled by the Kossakowski--Lindblad equation~\cite{KOSSAKOWSKI1972247,Lindblad:1975ef}, and it has been used by Susskind, Unruh, Hawking, and others, to characterize the evolution of pure states into mixed states in evaporation of black holes~\cite{Banks:1983by,Unruh:1995gn,Hawking:1976ra}. Although this effect is assumed to be very small, its accumulation along the Universe's history is recorded in an effective cosmological constant that drives the cosmic acceleration~\cite{Josset:2016vrq}. Even more, such a process provides a resolution to the Hubble tension~\cite{Perez:2020cwa} and it might explain the low spin of black holes detected via gravitational waves~\cite{Perez:2019gyd}. Diffusion may produce measurable effects in compact objects~\cite{Astorga-Moreno:2019uin,Alho:2016syc} and in cosmology~\cite{Calogero:2012kd,Calogero:2013zba,Benisty:2017eqh,Benisty:2017lmt,Benisty:2018oyy,Daouda:2018kuo} that can be constrained by current experimental data. 

In a homogeneous and isotropic Universe, energy diffusion could lead to third-order time derivatives of the scale factor~\cite{Garcia-Aspeitia:2019yni,Garcia-Aspeitia:2019yod}, introducing the jerk parameter of cosmography~\cite{Weinberg:2008zzc} into the Friedmann's equations. However, it is well known that higher-order time derivatives may produce undesirable features from an initial-value problem viewpoint, complicating its physical interpretation.  

In this work, we point out that the latter issue can be avoided by assuming an equation of state (EoS) that relates the diffusion function with the energy density of matter. This choice is inspired by the mass-proportional continuous spontaneous localization (CSL)~\cite{Pearle:1976ka,Ghirardi:1985mt,Pearle:1988uh,Ghirardi:1989cn}, where energy is created due to quantum collapse. To this end, we start from the modified conservation law in UG arising from the restricted diffeomorphism invariance. Then, we study a barotropic model that encompasses the physics behind the energy diffusion process and later we focus on the CSL model. In a matter dominated era, we solve the field equations analytically by assuming that only dark matter develops energy diffusion. Different accelerated solutions with phase transition are found. We contrast the distinct models with measurements of the Hubble parameter and receding velocity of SNe Ia. The observational limits allow for the barotropic model to be compatible with very small but nontrivial energy diffusion at the background level, driving the accelerated expansion without the cosmological constant problem. In the case of the CSL model, our analysis shows it can fit the data proficiently, however, either the best-fit parameters are not compatible with previous experimental data or it does not give a best fit when the latter is taken into account beforehand. This fact reinforces the result of Ref.~\cite{Martin:2019jye}, where similar conclusions were obtained from observations of the CMB. 

The article is organized as follows: in Sec.~\ref{sec:unimodular}, a review of unimodular gravity is given. In Sec.~\ref{sec:cosmo}, we present the two models under consideration and solve the field equations analytically for a homogeneous and isotropic Universe. In Sec.~\ref{sec:constraints}, we use the observational evidence coming from SNe Ia and observational Hubble data to constraint both models and in Sec.~\ref{sec:results} we present our main results. Finally, Sec.~\ref{sec:conclusions} is devoted to conclusions and further remarks. Throughout the manuscript, we use the metric signature $(-,+,+,+)$, the Riemann tensor is $R^{\lambda}{}_{\rho\mu\nu} = \partial_\mu \Gamma^{\lambda}{}_{\rho\nu} + ...$, while the Ricci tensor and scalar are defined as $R_{\mu\nu} = R^{\lambda}{}_{\mu\lambda\nu}$ and $R=g^{\mu\nu}R_{\mu\nu}$, respectively.

\section{Unimodular gravity\label{sec:unimodular}}

Unimodular gravity can be described in different but equivalent ways~\cite{Einstein:1919gv,Anderson:1971pn,vanderBij:1981ym,Buchmuller:1988wx,Unruh:1988in,Henneaux:1989zc,Ng:1990xz,Finkelstein:2000pg,Ellis:2010uc}. Here, we focus on those realizations given by an action principle that remains invariant under volume-preserving diffeomorphisms, i.e. those generated by vector fields $\xi^\mu$ satisfying $\nabla_\mu \xi^\mu = 0$. This can be done by introducing a Lagrange multiplier $\lambda(x)$ that fixes the volume element on shell according to (see~\cite{Padilla:2014yea})
\begin{align}\notag
 S\left[g_{\mu\nu},\lambda,\Psi\right] &= \frac{1}{2\kappa}\int\dv\left[R - 2\lambda(x)\left(1-\frac{\varepsilon_0(x)}{\sqrt{-g}} \right) \right]\\
  \label{ugaction}
  &\quad + \int\dv\,\Lag_m\left[g_{\mu\nu},\Psi \right],
\end{align}
where $\kappa=8\pi G_N$, $\Lag_m$ is the matter Lagrangian, $\Psi$ are the matter fields, and $\varepsilon_0(x)$ is a nondynamical $4$-form that breaks the diffeomorphism invariance down to volume-preserving diffeomorphisms. 

It is well known that diffeomorphism invariance of the matter action generated by an \emph{arbitrary} vector field imples the conservation law $\nabla^\mu T_{\mu\nu} = 0$. Therefore, the reduced symmetry of UG must modify the latter, since vector fields that generate the symmetries of~\eqref{ugaction} are no longer arbitrary but subject to the condition of being divergence free. In order to obtain the modified conservation law, the restriction $\nabla_\mu\xi^\mu=0$ can be solved as $\xi^\mu = \tfrac{1}{2}\epsilon^{\mu\nu\lambda\rho}\nabla_\lambda\alpha_{\mu\nu}$, where $\alpha_{\mu\nu}$ is an \emph{arbitrary} $2$-form. The Noether theorem associated to this reduced symmetry implies that $\nabla_{[\mu}\nabla^\lambda T_{\nu]\lambda} = 0$. This, in turn, can be solved locally through the Poincar\'e lemma to obtain~\cite{Josset:2016vrq} 
\begin{align}\label{noether}
 \nabla^\mu\left(T_{\mu\nu} - g_{\mu\nu}Q \right) = 0.
\end{align}
Here, $Q=Q(x)$ is an arbitrary function that measures the nonconservation of the energy-momentum tensor and hereafter is referred to as the diffusion function. In fact, if $Q=\mbox{constant}$, the usual conservation law is obtained. The case $Q\neq\mbox{constant}$ will play a key role in the forthcoming analysis.

The field equations are obtained by performing arbitrary variations of~\eqref{ugaction} with respect to $g_{\mu\nu}$, $\lambda$, and $\Psi$, giving
\begin{align}\label{eomg}
R_{\mu\nu} - \frac{1}{2}g_{\mu\nu} R + \lambda(x)g_{\mu\nu} &= \kappa T_{\mu\nu},\\
\label{eoml}
\sqrt{-g} &= \varepsilon_0,\\
\frac{\delta\Lag_m}{\delta\Psi} &= 0,
\end{align}
respectively. Equation~\eqref{eomg} can be interpreted as the Einstein's field equations with a cosmological function $\lambda(x)$, bearing in mind that the metric $g_{\mu\nu}$ has its volume element fixed through Eq.~\eqref{eoml}. Taking the trace on Eq.~\eqref{eomg}, the variable cosmological constant can be solved algebraically as $\lambda(x) = \left(\kappa T + R \right)/4$, with $T$ being the trace of the energy-momentum tensor. Replacing it back into~\eqref{eomg}, one obtains
\begin{align}\label{tracelesseom}
 R_{\mu\nu} - \frac{1}{4}g_{\mu\nu}R = \kappa\left(T_{\mu\nu} - \frac{1}{4}g_{\mu\nu}T \right),
\end{align}
which is identified as the traceless part of the Einstein's field equations. This fact shows that UG possesses one independent equation less than GR with an additional constraint [cf. Eq.~\eqref{eoml}]. Taking the covariant divergence on Eq.~\eqref{tracelesseom} and using the conservation law in Eq.~\eqref{noether}, a first integral of motion is obtained
\begin{align}\label{variablelambda}
 \lambda(x) &= \Lambda + \kappa Q(x) ,
\end{align}
where $\Lambda$ is an integration constant that should be fixed by initial data. Thus, it becomes clear that the variable cosmological constant is sourced by the energy diffusion function $Q(x)$. Interestingly, Eq.~\eqref{noether} allows for $\nabla_\mu\lambda(x)\neq0$ after Bianchi identities, in contrast to promoting the cosmological constant to be a field in the standard Einstein--Hilbert action. Moreover, the vacuum energy density of quantum fields does not gravitate, since the right-hand side of Eq.~\eqref{eomg} is invariant under the simultaneous shift symmetry $T_{\mu\nu} \to T_{\mu\nu} + \langle\rho\rangle g_{\mu\nu}$ and $Q\to Q + \langle\rho\rangle$. 

In the next section, we assume an homogeneous and isotropic Universe filled with matter, and find $Q(x)$ from the field equations in two different models.




\section{Cosmic acceleration from diffusion\label{sec:cosmo}}

Assuming that the Universe is homogeneous and isotropic at large scales, the line element is given by the Friedmann--Lema\^{i}tre--Robertson--Walker (FLRW) metric 
\begin{align}
 \diff{s^2} &= -\diff{t^2} + a^2(t)\left(\frac{\diff{r^2}}{1-kr^2}  + r^2\diff{\vartheta^2} + r^2\sin^2\vartheta\diff{\varphi^2}\right),
\end{align}
where $a(t)$ denotes the scale factor and $k=\pm1,0$ stands for spherical, hyperbolic, and flat spatial sections, respectively. Additionally, we consider a perfect fluid that describes the matter sector with an energy-momentum tensor given by
\begin{align}
T_{\mu\nu} = p\, g_{\mu\nu} + \left(\rho + p \right)u_\mu u_\nu,    
\end{align} 
where $p$, $\rho$ and $u_\mu$ are the pressure, energy density and fluid element four-velocity, respectively, normalized as $u_\mu u^\mu = -1$. Furthermore, we shall assume that $T_{\mu\nu}$ and the diffusion function are isotropic and homogeneous as well, implying that $T_{tt}=\rho(t)$, $T_{ij}=a^2(t)p(t)\delta_{ij}$, and $Q=Q(t)$.

For these ans\"atze, Eq.~\eqref{eomg} leads to the Friedmann's equations in UG, which are
\begin{align}\label{friedmann1}
 3H^2 &= \kappa\sum_i\left(\rho_i + Q_i \right) + \Lambda - \frac{3k}{a^2},\\
 \label{friedmann2}
 2\dot{H} + 3H^2 &= - \kappa\sum_i\left(p_i-Q_i \right) + \Lambda - \frac{k}{a^2},
\end{align}
where dot denotes derivative with respect to the cosmic time, $H(t)=\dot{a}/a$ is the Hubble function, and the sum over all species is assumed. Notice that we have considered different $Q$'s for the sake of generality. The constraint Eq.~\eqref{eoml}, on the other hand, reads
\begin{align}\label{varepsflrw}
 \varepsilon_0 = \frac{a^3 r^2 \sin\vartheta}{\sqrt{1-kr^2}},
\end{align}
while the conservation law for matter in Eq.~\eqref{noether} becomes
\begin{align}\label{conservation}
\dot{\rho}_i + \dot{Q}_i + 3H\left(\rho_i+p_i \right) = 0. 
\end{align}

The energy diffusion function, and therefore the variable cosmological constant, can be modeled in different ways. For this purpose, we consider two distinct scenarios in the next Subsections. The first one, is a barotropic model that encompasses the phenomenon we are interested in, while the second one is a well-studied scenario known as the CSL model. This will provide a setup to understand the dynamical evolution of $\lambda(t)$ in different contexts.

\subsection{A barotropic model\label{sec:toymodel}}

We propose a model where the pressure and diffusion function are parametrized by a barotropic EoS, i.e. $p=w\rho$ and $Q=x\rho$, with $w$ and $x$ being real constants. Physically, this means that diffusion is proportional to the derivative of energy density with respect to the cosmic time and its proportionality constant, $x$, is expected to be small. In such a case, Eq.~\eqref{conservation} can be solved as
    \begin{align}\label{rhosoltypeI}
 \rho_i = \rho_{0i}\, a^{-\frac{3\left(w_i+1\right)}{x_i+1}},
\end{align}
where $x_i\neq-1$ and $\rho_{0i}$ is an integration constant.\footnote{Imposing $x=1$ before the integration implies that $\rho=-p$, which represents the EoS for dark energy. Hereafter, we restrict ourselves to the case where $x_i\neq-1$.} Additionally, in order for the energy density to scale as the inverse of some power of the scale factor, the condition $x_i+1>0$ must be met for $w_i>-1$. We assume that this restriction holds from hereon.

Defining the deceleration parameter $q=-\ddot{a}/(H^2 a)$, the Friedmann's equations can be written in terms of the redshift $z$ as
\begin{align}\notag
    \frac{H(z)^2}{H_0^2} &= \sum_i (1+x_i)\Omega_{0i}(1+z)^{\frac{3(w_i+1)}{x_i+1}} \\
    \label{f1z}
    &\quad + \Omega_{0\Lambda} + \Omega_{0k}\left(1+z \right)^2 , \\
\label{qdef}
    \frac{H(z)^2 q(z)}{H_0^2} &= \frac{1}{2}\sum_i\left(1+3w_i - 2x_i \right)\Omega_{0i}\left(1+z\right)^{\frac{3(w_i+1)}{x_i+1}} - \Omega_{0\Lambda}.
\end{align}
Here, we have defined $\Omega_{i} = \kappa\rho_{i}/(3H^2)$, $\Omega_{\Lambda} = \Lambda/(3H^2)$, and $\Omega_{k} = -k/H^2$ as the dimensionless density parameters, while the subscript zero denotes evaluation at present time, e.g. $H_0=H(0)$, $\Omega_{0i} = \kappa\rho_{0i}/(3H_0^2)$, and so on.  Throughout this manuscript, we use the normalization $a(0)=1$. In general, each diffusion function admits a different EoS, namely $Q_i = x_i\rho_i$ (no sum over $i$).

Since $\Lambda$ is an integration constant, it should be fixed by initial data. To do so, we take the first integral of motion~\eqref{variablelambda} and evaluate it in the FLRW ansatz, giving
\begin{align}\label{firsintegral}
\dot{H} + 2H^2 &= \frac{\kappa}{6}\sum_i\left(\rho_i- 3p_i +4Q_i \right)+\frac{2\Lambda}{3} - \frac{k}{a^2}.
\end{align}
It is straightforward to show that this expression is not independent of the Friedmann equations~\eqref{friedmann1} and~\eqref{friedmann2}. The value of the integration constant can be found by evaluating Eq.~\eqref{firsintegral} at present time, giving
\begin{align}\label{lambdavalue}
 \frac{2\Lambda}{3} &= H_0^2\left(1-q_0 \right) + k + \frac{H_0^2}{2}\sum_i\left(3w_i - 4x_i - 1 \right)\Omega_{0i},
\end{align}
where $q_0$ denotes the value of the deceleration parameter at $t=0$. One can use the second Friedmann equation to simplify this expression further. Evaluating Eq.~\eqref{qdef} at present time and replacing the value of $q_0$ in the last expression, we get
\begin{align}\label{omlambvalue}
 \Omega_{0\Lambda} &= 1 - \Omega_{0k} - \sum_i\left(1+x_i \right)\Omega_{0i},
\end{align}
which can be obtained analogously by evaluating Eq.~\eqref{f1z} at present time. Therefore, Eq.~\eqref{omlambvalue} represents the Friedmann's constraint at present time in the presence of energy diffusion, which implies that $\Omega_{0\Lambda}$ is completely determined by $\Omega_{0k}$, $x_i$, $\Omega_{0i}$, and $H_0$.  Nevertheless, since the parameter $x_i$ is assumed to be constant during the cosmic evolution, its present value must be determined by cosmological data. In this sense, instead of having two free parameters as in the $\Lambda$CDM model after the Friedmann's constraint, i.e. $\Omega_{0k}$ and $\Omega_{0m}$, this model has three. 

\begin{figure}
      \centering
      \includegraphics[scale=0.45]{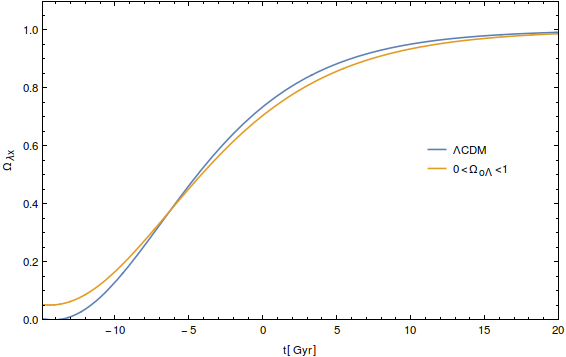}
      \caption{Cosmic evolution of the dimensionless parameter $\Omega_{\lambda x}$ associated to the variable cosmological constant $\lambda(t)$ in the barotropic model with $0<\Omega_{0\Lambda}<1$.}
      \label{toy-densities}
  \end{figure}

In a flat FLRW Universe dominated by dust, the energy density scales as $\rho_m=\rho_{0m}a^{-3}$ and the value of the dimensionless density associated to the integration constant, $\Omega_\Lambda$, obeys
\begin{align}\label{Friedmannconstraint}
  1  &= \Omega_m + \left(\Omega_{\Lambda}  +x \Omega_{m}\right) \equiv \Omega_m + \Omega_{\lambda x},
\end{align}
where $\Omega_{m} = \kappa\rho_{m}/(3H^2)$ and $\Omega_{\lambda x}$ is the dimensionless parameter associated to the variable cosmological constant $\lambda(t)$ in the barotropic model. Thus, the number of free parameters in this case is reduced by one. The cosmic evolution of $\Omega_{\lambda x}$ is given in Fig.~\ref{toy-densities} for the unique solution having phase transition in this model.\footnote{The existence of future singularities induced by diffusion as shown in forthcoming sections implies that the redshift is not a monotonically decreasing function of time. Thus, hereon we use the cosmic time for the plots instead of the redshift.}

In order to avoid negative energy density for matter fields, the condition $\Omega_{m}\geq0$ must be met. This, in turn, implies 
\begin{align}\label{condlamb}
    \frac{1-\Omega_{\Lambda}}{1+x} \geq 0.
\end{align}
Since we have demanded $1+x>0$ for a reasonable scaling of the matter's energy density, condition~\eqref{condlamb} implies that $\Omega_{\Lambda}\leq1$. This bound is saturated when $\Omega_{\Lambda}=1$, which is equivalent to demanding $\Omega_{m}=0$ for $x\neq1$. 

Then, the Friedmann's equations in a flat Universe for a matter dominated era are
\begin{align}
    \frac{H(z)^2}{H_0^2} &= (1+x)\Omega_{0m}(1+z)^{\frac{3}{x+1}} + \Omega_{0\Lambda} , \label{HToyModel} \\
    \frac{H(z)^2 q(z)}{H_0^2} &= \frac{1}{2}\left(1- 2x \right)\Omega_{0m}\left(1+z\right)^{\frac{3}{x+1}} - \Omega_{0\Lambda}.
\end{align}
These equations admit exact solutions in four distinctive cases: (i)~$\Omega_{0\Lambda} = 0$, (ii)~$\Omega_{0\Lambda} = 1$, (iii)~$0<\Omega_{0\Lambda}<1$, and (iv)~$\Omega_{0\Lambda}<0$, which should be treated separately. In what follows, we define the parameter $y\equiv -3/(1+x)$ where, due to the previous considerations, implies that $y<0$.

\subsubsection{Case when $\Omega_{0\Lambda} = 0$}

This case corresponds to a Universe filled only with a dark matter, whose energy density is modified by the factor $(1+x)$ introduced by $Q(x)$ [cf. Eq.~\eqref{Friedmannconstraint}]. As we will see below, the solution for the scale factor resembles the one obtained for the Einstein--de Sitter model, i.e. 
    \begin{align}
        a(t) &= \left[\frac{\sqrt{-3y\Omega_{0m}}}{2}H_0(t-t_0) \right]^{-\frac{2}{y}},
    \end{align}
    where $t_0$ is an integration constant. Since $y<0$, this solution represents a power-law expansion without phase transition, and therefore is disfavored by observations. It can be decelerated or accelerated for if $-2/y$ is lesser or greater than 1, respectively. 

\subsubsection{Case when $\Omega_{0\Lambda} = 1$}

Since we have previously demanded $1+x>0$, this case implies $\Omega_{0m}=0$ from   Eq.~\eqref{Friedmannconstraint}. Thus, the evolution corresponds to a de Sitter solution with a scale factor 
    \begin{align}
      a(t) &= a_0 e^{H_0 t},
  \end{align}
 where $a_0$ is an integration constant and $H_0 = \sqrt{\Lambda /3}$. We shall not consider this solution from now on. 
 
\subsubsection{Case when $0<\Omega_{0\Lambda} <1$}

\begin{figure}
      \centering
      \includegraphics[scale=0.39]{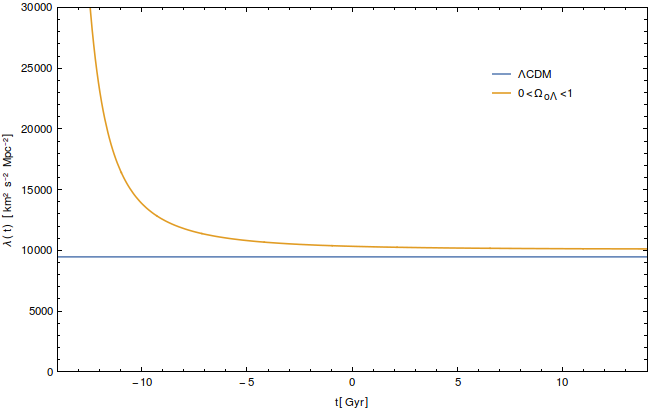}
      \caption{Behavior of $\lambda(t)$ defined in Eq.~\eqref{variablelambda} for the barotropic model when $0<\Omega_{0\Lambda}<1$, considering the best-fit values of Table~\ref{bestfittable}. For comparison, we include the values of $\Lambda$CDM with $\Lambda=3H_0^2 \Omega_{0\Lambda}$, where the mean values $\Omega_{0\Lambda}=0.689$ and $H_0=67.66\,\si{km.s^{-1}.Mpc^{-1}}$ have been used~\cite{Aghanim:2018eyx}.}
      \label{toy-variable-lambda}
  \end{figure}

In this scenario, dark matter coexists with negative pressure dark energy modeled by the variable cosmological constant $ \lambda(t) = \Lambda + \kappa Q(t)$. Since the dark matter density is a decreasing function of time, the variable cosmological constant also decreases as the cosmic time evolves, going to a constant value $\Lambda$ (see Fig.~\ref{toy-variable-lambda}).\footnote{A similar behavior is found for the CSL model of Sec.~\ref{sec:CSL}, that has been previously outlined in Ref.~\cite{Josset:2016vrq}.} Therefore, it is expected that a phase transition from a decelerated to accelerated expansion occurs, ending with a de Sitter-like behavior. In fact, the Friedmann's equations in this case are solved by
  \begin{align}\label{toysolsinh}
      a(t) &= \left(-\frac{y\Omega_{0\Lambda}}{3\Omega_{0m}} \right)^{\frac{1}{y}}\sinh^{-\frac{2}{y}}\left[-\frac{y\sqrt{\Omega_{0\Lambda}}}{2}H_0(t-t_0) \right].
  \end{align}
 Thus, it becomes clear that this solution possesses a phase transition and its behavior is more likely to describe the observational data.  The age of the Universe in this case is
  \begin{align}\label{agetoy}
      \tau = -\frac{2}{yH_0\sqrt{\Omega_{0\Lambda}}}\arcsinh\left(\sqrt{-\frac{y\Omega_{0\Lambda}}{3\Omega_{0m}}}\right),
  \end{align}
  where the integration constant $t_0$ has been fixed according to the normalization $a(0)=1$. This solution will be taken into account in forthcoming sections to be constrained with cosmological data.

\subsubsection{Case when $\Omega_{0\Lambda}<0$}

The last possibility characterizes a Universe filled with dark matter and a negative integration constant $\Lambda$. This solution represents an anti-de Sitter Universe with future cosmological singularity, whose scale factor is given by
  \begin{align}
    a(t) &= \left(\frac{y\Omega_{0\Lambda}}{3\Omega_{0m}} \right)^{\frac{1}{y}} \cos^{-\frac{2}{y}}\left[\frac{y\sqrt{-\Omega_{0\Lambda}}}{2} H_0(t-t_0) \right],
  \end{align}
where, recall, $y<0$. However, it does not exhibit a phase transition and it is therefore unsupported by experimental data.

In sum, the barotropic model develops energy diffusion through dark matter where, for positive values of $x$, it implies a energy loss as $Q(t)$ evolves. Four different exact solutions are found, depending on the value of integration cosmological constant. From an observational viewpoint, the most interesting one appears when $0<\Omega_{0\Lambda}<1$, since it has a phase transition. This solution will be contrasted with SNe Ia and observational Hubble data later.

In the next section, we consider the specific form of the diffusion function obtained in Ref.~\cite{Josset:2016vrq} for the CSL model in a cosmological setup and solve the field equations analytically.  

\subsection{Continuous spontaneous localization\label{sec:CSL}}

The CSL model~\cite{Pearle:1976ka,Ghirardi:1985mt,Pearle:1988uh,Ghirardi:1989cn,Bassi:2003gd} is one of the most studied and refined approaches to solve the measurement problem of quantum mechanics. It describes quantum-to-classical transitions through stochastic and nonlinear terms in the Schr\"odinger equation, that produce the quantum collapse of the wave function. Remarkably, the appearance of energy divergences in collapse models~\cite{PhysRevA.43.9,Pearle:1995qp,Bassi_2005} can be avoided by introducing energy dissipation of matter~\cite{PhysRevA.90.062135,Smirne-Bassi}, which has been shown to be compatible with the volume-preserving diffeomorphisms invariance of UG~\cite{Josset:2016vrq}.\footnote{It is worth mentioning that an additional framework with similar predictions is the causal set approach to quantum gravity~\cite{Dowker:2003hb,Philpott:2008vd}. For concreteness, we restrict ourselves to the CSL setup and leave causal sets for a future studies.} In this Subsection, we focus on this scenario that provides a specific prediction for the diffusion function $Q(t)$.

For matter, the CSL model predicts a characteristic form of the diffusion function that can be expressed as~\cite{Josset:2016vrq}\footnote{In Ref.~\cite{Josset:2016vrq}, this form was taken for baryons only. Here, we consider that dark matter behaves similarly to baryons, although an extension that distinguishes their nature is straightforward.}
\begin{align}\label{diffusioncsl}
    Q_{\rm CSL}(t) &= Q_0 - \xi_{\rm CSL}\int\rho_m\diff{t},
\end{align}
where $Q_0$ is an integration constant and $\rho_m$ is the energy density of the matter content. As we will see next, $Q_0$ contributes to an effective value of the integration constant that plays the role of the cosmological constant. The latter will be fixed by initial data through their corresponding Friedmann's constraint. Moreover, current experimental data impose severe constraints on the free parameter according to $3.3\times 10^{-42}\,\mbox{s}^{-1}<\xi_{\rm CSL}<2.8\times 10^{-29}\,\mbox{s}^{-1}$  (see~\cite{Josset:2016vrq} and references therein). 

Assuming a dustlike dark matter component, the conservation law~\eqref{conservation} is solved by 
\begin{align}\label{rhocsl}
\rho_m &= \rho_{0m} e^{\xi_{\rm CSL}t}\,a^{-3}, 
\end{align}
where $\rho_{0m}$ is an integration constant related to the present energy density of matter. Thus, the effect of energy diffusion in the CSL model is to provide a modulation of the scaling of matter. Although one would expect an exponential grow of energy density at late times, the backreaction of diffusion generates a similar modification on the scale factor [see Eqs.~\eqref{asolcsl1}-\eqref{asolcsl3} below] that cancels out the exponential in Eq.~\eqref{rhocsl}, producing a decay on the energy density as long as the scale factor grows.

Considering that only dark matter develops diffusion, the Friedmann's equations in the CSL scenario become
\begin{align}\notag
\frac{H(z)^2}{H_0^2} &= \left[e^{\xi_{\rm CSL}t}(1+z)^3 - \xi_{\rm CSL}\int e^{\xi_{\rm CSL}t}(1+z)^3\diff{t} \right]\Omega_{0m} \\ 
\label{friedmann1csl}
&\quad + \Omega_{0\Lambda_{\rm eff}} + \Omega_{0k}(1+z)^2, \\
\notag
\frac{q(z)H(z)^2}{H_0^2} &= \left[\frac{e^{\xi_{\rm CSL}t}(1+z)^3}{2} + x\int e^{\xi_{\rm CSL}t}(1+z)^3\diff{t} \right]\Omega_{0m} \\
\label{friedmann2csl}
&\quad - \Omega_{0\Lambda_{\rm eff}},
\end{align}
where $\Omega_{\Lambda_{\rm eff}}=\Omega_{\Lambda} + \kappa Q_0/(3H^2)$ and $\Omega_{0\Lambda_{\rm eff}}$ represents its value at present time. In the flat case, and defining the quantity $\Omega_{\xi} \equiv \xi_{\rm CSL}^2/(9H^2)>0$ (with $\Omega_{0\xi}$ denoting its present value), we find that these equations admit exact solutions in three distinct cases: (i) $\Omega_{0\Lambda_{\rm eff}} > \Omega_{0\xi} $, (ii) $\Omega_{0\Lambda_{\rm eff}} < \Omega_{0\xi} $, and (iii) $\Omega_{0\Lambda_{\rm eff}} = \Omega_{0\xi}$. 

The Friedmann's constraints for each case are
\begin{align}\label{FriedmannconstraintCSL1}
1 &= \sqrt{\Omega_\xi} \pm \sqrt{\Omega_m + \Omega_{\Lambda_{\rm eff}} -\Omega_\xi},\\
\label{FriedmannconstraintCSL3}
1 &= \sqrt{\Omega_\xi} + \sqrt{\Omega_m},
\end{align}
where the upper and lower signs of Eq.~\eqref{FriedmannconstraintCSL1} correspond to the first and second cases, respectively, while Eq.~\eqref{FriedmannconstraintCSL3} corresponds to the third case. To obtain a clearer meaning of Eqs.~\eqref{FriedmannconstraintCSL1} and~\eqref{FriedmannconstraintCSL3} 
before the analytical solutions are presented, we subtract $\sqrt{\Omega_\xi}$ in their both sides and take their second power. Both results can be summarized in  \begin{align}\label{FriedmannconstraintCSL10}
1 &= \Omega_m + \Omega_{\Lambda_{\rm eff}} + 2\sqrt{\Omega_\xi} - 2\Omega_\xi \equiv \Omega_m + \Omega_{\lambda\xi},
\end{align}
where $\Omega_{0\Lambda_{\rm eff}} = \Omega_{0\xi}$ is contained as a particular case. Thus, $\Omega_{\lambda\xi}$ is interpreted as the dimensionless parameter associated to the variable cosmological constant $\lambda(t)$ for the CSL model. Remarkably, the case when  $\Omega_{0\Lambda_{\rm eff}} < \Omega_{0\xi}$ [i.e. with the minus sign in Eq.~\eqref{FriedmannconstraintCSL10}], admits a sign flip of the variable cosmological constant as time evolves, passing from a de Sitter to an anti de Sitter phase at very late times. Its cosmic evolution is displayed in Fig.~\ref{csl-densities} for the three analytical solutions found below.

In the following, we discuss each of these cases separately and provide the analytic form of the scale factor that solves Eqs.~\eqref{friedmann1csl} and~\eqref{friedmann2csl}. Remarkably, it is found that the presence of energy diffusion generates an exponential modulation of the scale factor in all cases. This feature induces a phase transition from decelerated to accelerated expansion, providing a suitable setup to be constrained by cosmological data.

\subsubsection{Case when $\Omega_{0\Lambda_{\rm eff}} > \Omega_{0\xi} $}

This scenario represents a Universe filled with dark matter, where the effective integration constant dominates over energy diffusion at present time. Therefore, in the constraint given by Eq.~\eqref{FriedmannconstraintCSL10}, the variable cosmological constant remains positive during the whole evolution. In fact, solving the Friedmann's equations~\eqref{friedmann1csl} and~\eqref{friedmann2csl}, we obtain a scale factor 
\begin{align}\notag
    a(t) &= \left(\frac{\Omega_{0m}}{\Omega_{0\Lambda_{\rm eff}} - \Omega_{0\xi}} \right)^{\frac{1}{3}}\exp\left[\sqrt{\Omega_{0\xi}}H_0 t\right]\\
    \label{asolcsl1}
   &\quad \times \sinh^{\frac{2}{3}}\left[\frac{3\sqrt{\Omega_{0\Lambda_{\rm eff}} - \Omega_{0\xi}}}{2}H_0\left(t-t_0 \right) \right].
\end{align}
Remarkably, this solution represents a Universe with phase transition from deceleration to acceleration driven by energy diffusion induced by quantum collapse, with a de Sitter-like expansion at late times. The age of the Universe in this scenario is
\begin{align}
\tau = \frac{2}{3H_0\sqrt{\Omega_{0\Lambda_{\rm eff}} - \Omega_{0\xi}}} \arcsinh\left(\sqrt{\frac{\Omega_{0\Lambda_{\rm eff}} - \Omega_{0\xi}}{\Omega_{0m}}}\right),
\end{align}
where the integration constant $t_0$ has been fixed by $a(0)=1$.

\subsubsection{Case when $\Omega_{0\Lambda_{\rm eff}} < \Omega_{0\xi} $}

\begin{figure}
    \centering
    \includegraphics[scale=0.34]{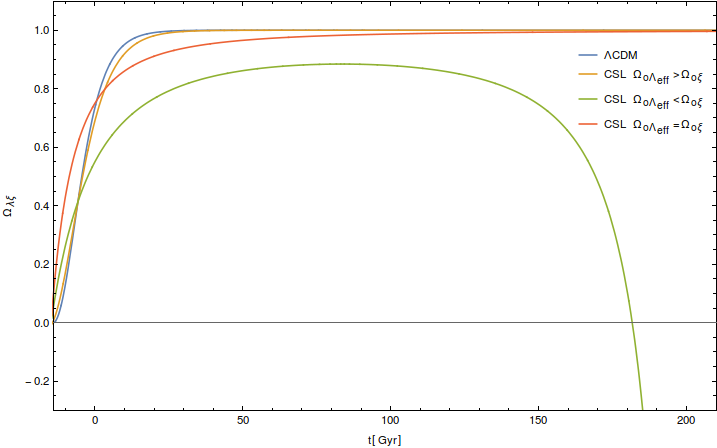}
    \caption{Cosmic evolution of the dimensionless parameter $\Omega_{\lambda\xi}$ associated to the variable cosmological constant $\lambda(t)$ in the CSL model.}
    \label{csl-densities}
\end{figure}

\begin{figure}
\centering
    \includegraphics[scale=0.39]{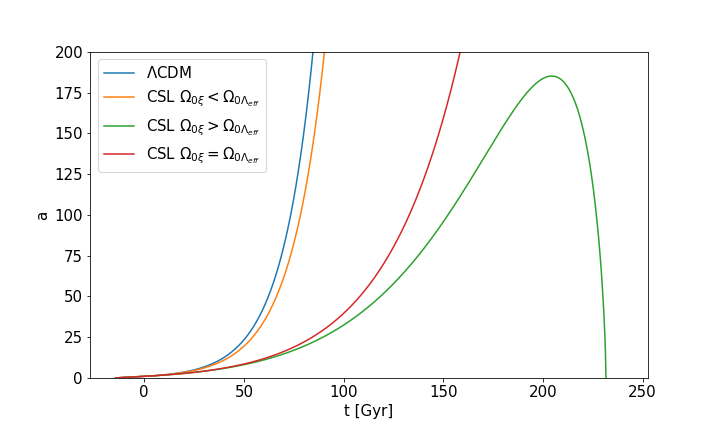}
    \caption{Very far future behavior of scale factors of CSL and $\Lambda$CDM models, considering the best-fit values of Table~\ref{bestfittable}.}
    \label{scalefactorcslfuture}
\end{figure}

\begin{figure}
    \centering
    \includegraphics[scale=0.44]{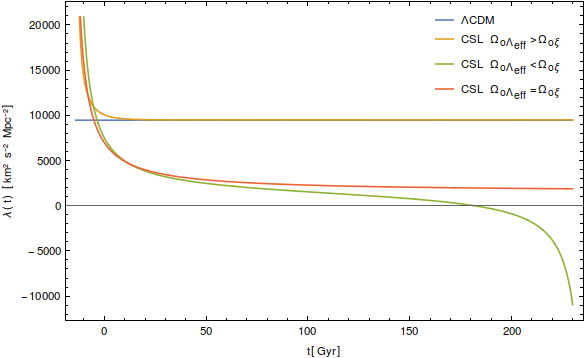}
    \caption{behavior of $\lambda(t)$ defined in Eq.~\eqref{variablelambda} for the three solutions of the CSL model, using the best-fit values of Table~\ref{bestfittable}. The case of $\Lambda$CDM is included for the sake of comparison, where we use $\Lambda=3H_0^2 \Omega_{0\Lambda}$ with mean values $\Omega_{0\Lambda}=0.689$ and $H_0=67.66\,\si{km.s^{-1}.Mpc^{-1}}$ as reported in Ref.~\cite{Aghanim:2018eyx}.}
    \label{csl-variable-lambda}
\end{figure}

In this scenario, the Universe is filled with dark matter whose diffusion function dominates over the cosmological integration constant. The constraint given by Eq.~\eqref{FriedmannconstraintCSL1} indicates that $\Omega_m$ decreases to a minimum value and the variable cosmological constant takes negative values at late times (see Figs.~\ref{csl-densities} and~\ref{csl-variable-lambda}). This behavior is encoded in the solution of the scale factor as a function of the cosmic time, i.e.,
\begin{align}\notag
    a(t) &= \left(\frac{\Omega_{0m}}{\Omega_{0\xi}- \Omega_{0\Lambda_{\rm eff}}  } \right)^{\frac{1}{3}}\exp\left[\sqrt{\Omega_{0\xi}}H_0 t\right] \\
    \label{asolcsl2}
  &\quad  \times \cos^{\frac{2}{3}}\left[\frac{3\sqrt{\Omega_{0\xi}- \Omega_{0\Lambda_{\rm eff}} }}{2}H_0\left(t-t_0 \right) \right].
\end{align}
Therefore, the exponential modulation induced by quantum collapse generates a phase transition during the period when the variable cosmological constant remains positive. As time evolves, it can take negative values, leading to a big crunch for very far future times. The behavior of the scale factor of this solution is plotted in Fig.~\ref{scalefactorcslfuture}, while the variable cosmological constant is included in Fig.~\ref{csl-variable-lambda}. The age of the Universe in this case is given by  
  \begin{align}
  \tau &= \frac{2}{3H_0\sqrt{\Omega_{0\xi}-\Omega_{0\Lambda_{\rm eff}}}}\left[\frac{\pi}{2} + \arccos\left(\sqrt{\frac{\Omega_{0\xi} - \Omega_{0\Lambda_{\rm eff}}}{\Omega_{0m}}}\right) \right],
 \end{align}
where, as before, the integration constant $t_0$ has been fixed according to $a(0)=1$.

\subsubsection{Case when $\Omega_{0\Lambda_{\rm eff}} = \Omega_{0\xi} $}

Whenever the dimensionless parameters related to the effective integration constant and energy diffusion are equal at present time, the function $\Omega_\xi$ goes to its minimum value as long as $\Omega_m$ goes to zero with the cosmic expansion. Thus, the variable cosmological constant reaches a nontrivial asymptotic value and it is expected to have a de Sitter-like expansion at late times driven by $\Omega_\xi$. Integrating Eqs.~\eqref{friedmann1csl} and~\eqref{friedmann2csl} we obtain the scale factor, which takes the form 
\begin{align}\label{asolcsl3}
a(t) &= \left[\frac{3\sqrt{\Omega_{0m}}}{2}H_0\left(t-t_0\right) \right]^{\frac{2}{3}}\exp\left[\sqrt{\Omega_{0\xi}}H_0 t\right].
\end{align}
This solution represents a power-law expansion modulated by an exponential function of the cosmic time, with a de Sitter behavior at late times. Remarkably, it has a phase transition from decelerated to accelerated expansion due to the nontrivial contribution of energy diffusion. The age of the Universe in this case is 
\begin{align}
    \tau &= \frac{2}{3H_0\sqrt{\Omega_{0m}}},
\end{align}
where the normalization $a(0)=1$ has been used.

In the next section, we constrain the parameters of the barotropic and CSL models using observational data for a flat FLRW Universe. To this end, we focus on solutions possessing phase transition and compare them with observational Hubble data (OHD) and SNe Ia data.

\section{Cosmological constraints\label{sec:constraints}}

To impose the constraints on the preceding models, we compute the best-fit value of their parameters by means of the affine-invariant Markov Chain Monte Carlo method (MCMC)~\cite{Goodman_Ensemble_2010}. This is implemented in the pure-Python code \textit{emcee}~\cite{Foreman_emcee_2013} by setting 24 chains with 4500 steps. In order for the latter to explore the whole parameter space and get settled in the maximum of the probability density, 1500 burn-in steps are performed beforehand.

Since we are implementing Bayesian statistical analysis to estimate the free parameters and their confidence regions, we construct the Gaussian likelihood function
\begin{align}\label{distributions}
    \mathcal{L}_I = N \exp\left(-\frac{\chi^2_I}{2} \right).
\end{align}
Here, $N$ is a normalization constant and $I$ stands for each data set under consideration, namely OHD, SNe Ia, and their joint analysis with $\chi^{2}_{\rm joint}=\chi^{2}_{\rm OHD}+\chi^{2}_{\rm SNe}$. They are described in the following, alongside their $\chi^{2}$ functions.

\subsection{Observational Hubble Data}

To constrain the models with OHD, we use the data set of Ref.~\cite{Magana_Cardassian_2018} that consists of 51 Hubble data points in the redshift range $0.07\leq z\leq 2.36$. In this catalog, 31 data points are obtained by the differential age method~\cite{Jimenez_COnstraining_2002}, where the points can be estimated using
\begin{equation}
H(z)=-\frac{1}{1+z}\dfrac{\diff{z}}{\diff{t}},\label{DA}
\end{equation}
with $\diff{z}/\diff{t}$ being measured using the $4000 \si{\angstrom}$ break feature as a function of the redshift. This method is model independent since it relies on the metallicity and the age of the stellar population of early galaxies estimated through spectroscopy. The remaining data points come from measurements of baryon acoustic oscillation, assuming that the data of $H(z)$ is obtained independently. Hence, the merit function of the OHD is constructed as
\begin{equation}
\chi^{2}_{\rm OHD}=\sum_{i=1}^{51}{\left[\frac{H_{i}-H_{\rm th}\left(z_{i},\theta\right)}{\sigma_{H,i}}\right]^{2}}, \label{OHD}
\end{equation}
where $H_i$ and $H_{\rm th}$ are the observational and theoretical Hubble parameters at redshift $z_i$, respectively, $\sigma_{H,i}$ is the associated error of $H_i$, and $\theta$ denotes the free parameters.

It is worth mentioning that we consider the present value of the Hubble constant $H_0$ as a free parameter as well. Thus, for the Bayesian analysis, we use a Gaussian prior $G(\gamma,\delta)$ over the dimensionless Hubble parameter, $h$, where $\gamma$ and $\delta$ are the mean value and standard deviation, respectively. This is done according to $H_{0}=73.24\pm 1.74~\si{km.s^{-1}.Mpc^{-1}}$, measured with a $2.4\%$ of uncertainty~\cite{Riess_Determination_2016}.

\subsection{Type Ia Supernovae}

In addition to OHD, SNe Ia data can be used to constrain the models. Currently, there are three main catalogs: Union 2.1 with 557 points in the redshift range $0.015\leq z\leq 1.4$~\cite{Suzuki_Hubble_2012}, JLA with 740 points in the redshift range $0.01\leq z\leq 1.3$~\cite{Betoule_Improved_2014}, and Pantheon with 1048 point in the redshift range $0.01\leq z\leq 2.3$~\cite{Scolnic_Complete_2018}. Here we focus on the latter that is a compilation of 279 SNe Ia data discovered by the Pan-STARRS1 Medium Deep Survey, combined with their estimated distance from the Sloan Digital Sky Survey, Supernova Legacy Survey, and various low-$z$ and Hubble Space Telescope samples. In this case, the merit function is constructed as
\begin{equation} \label{SNeIa}
\chi^{2}_{\rm SNe}=\sum_{i=1}^{1048}{\left[\frac{\mu_{i}-\mu_{\rm th}\left(z_{i},\theta\right)}{\sigma_{i}}\right]^{2}},
\end{equation}
where $\mu_i$ and $\mu_{\rm th}$ are the observational and theoretical distance modulus of each SNe Ia at redshift $z_i$, respectively, $\sigma_i$ is the error in the measurement of $\mu_i$, and $\theta$ encompasses the free parameters of the respective model.

In the Pantheon sample, the observational distance modulus is obtained by using the modified version of the Tripp formula~\cite{Tripp_Luminosity_1998}. This method, however, is endowed with at least three nuisance parameters that must be jointly estimated with the cosmological parameters $\theta$. To overcome this problem, the method BEAMS with bias correction was proposed~\cite{Kessler_Correcting_2017}, in which its value reduces to
\begin{equation}
\mu_i=m_B-\mathcal{M}, \label{BBC}
\end{equation}
where $m_B$ is the apparent B-band magnitude of a fiducial SNe Ia and $\mathcal{M}$ is a nuisance parameter. Even more, the Pantheon sample gives the corrected apparent magnitude $m_B$ directly. On the other hand, the theoretical distance modulus in a flat FLRW spacetime of a given model is defined through the relation 
\begin{equation}
\mu_{\rm th}\left(z_{i},\theta\right)=5\log_{10}{\left[\frac{d_{L}\left(z_{i},\theta\right)}{\si{Mpc}}\right]}+\bar{\mu}, \label{mutheoretical}
\end{equation}
where $\bar{\mu}=5\left[\log_{10}{\left(c\right)}+5\right]$, $c$ is the speed of light given in units of $\si{km.s^{-1}}$, and $d_{L}$ is the luminosity given by
\begin{equation}
d_{L}\left(z_{i},\theta\right)=\left(1+z_{i}\right)\int_{0}^{z_{i}}{\frac{dz'}{H\left(z',\theta\right)}}. \label{lumdistance}
\end{equation}

Equation~\eqref{SNeIa} can be written in matrix notation (denoted by bold symbols) according to
\begin{equation}\label{SNeIamatrix}
\chi^{2}_{\rm SNe}=\mathbf{M}^\dagger\mathbf{C}^{-1}\mathbf{M},
\end{equation}
where $\mathbf{C}$ is the total covariance matrix given by
\begin{equation}\label{covariancematrix}
\textbf{C}=\textbf{D}_{\rm stat}+\textbf{C}_{\rm sys}, 
\end{equation}
and we have defined $\mathbf{M}=\mathbf{m}_{B}-\mbox{\boldmath$\mu$}_{\rm th}\left(z_{i},\theta\right)-\mbox{\boldmath$\mathcal{M}$}$. The entries of the diagonal matrix $\textbf{D}_{\rm stat}$ denote the statistical uncertainties of $m_B$ for each redshift, while $\textbf{C}_{\rm sys}$ denotes the systematic uncertainties in the BEAMS with bias correction approach.\footnote{The Pantheon data set is available online in the GitHub repository \href{https://github.com/dscolnic/Pantheon}{https://github.com/dscolnic/Pantheon}. The corrected apparent magnitude $m_{B}$ for each SNe Ia together with their respective redshifts and errors are available in the document \textit{lcparam\_full\_long.txt}. The full systematic uncertainties matrix is available in the document \textit{sys\_full\_long.txt}.} In order to reduce the number of free parameters and marginalize over $\mathcal{M}$, we define $\mathcal{M}=\bar{\mathcal{M}}-\bar{\mu}$ with $\bar{\mathcal{M}}$ being an auxiliary nuisance parameter and $\bar{\mu}$ being defined below Eq.~\eqref{mutheoretical}. Thus, Eq.~\eqref{SNeIamatrix} can be expanded as~\cite{Lazkos_Exploring_2005}
\begin{equation}\label{chi2projected}
\chi^{2}_{\rm SNe}=A\left(z,\theta\right)-2B\left(z,\theta\right)\bar{\mathcal{M}}+C\bar{\mathcal{M}}^{2}, 
\end{equation}
where
\begin{align}
A\left(z,\theta\right) & =  \bar{\mathbf{M}}^{\dagger}\textbf{C}^{-1}\bar{\mathbf{M}}, \\
B\left(z,\theta\right) & =  \bar{\mathbf{M}}^{\dagger}\textbf{C}^{-1}\,\textbf{1}, \\
C & =  \textbf{1}^{\dagger}\, \textbf{C}^{-1}\, \textbf{1}, \label{definitionsprojected}
\end{align}
with $\bar{\mathbf{M}}=\mathbf{m}_{B}-\mbox{\boldmath$\mu$}_{\rm th}\left(z_{i},\theta\right)+\bar{\mbox{\boldmath$\mu$}}$.


Minimizing Eq.~\eqref{chi2projected} with respect to $\bar{\mathcal{M}}$ gives $\bar{\mathcal{M}}=B/C$ and it reduces to
\begin{equation}
\chi^{2}_{\rm SNe}\Big|_{\rm min}=A\left(z,\theta\right)-\frac{B\left(z,\theta\right)^{2}}{C}. \label{SNeIaprojected}
\end{equation}
Notice that this function depends only on the free parameters of the model. In fact, Eq.~\eqref{SNeIamatrix} provides the same information as Eq.~\eqref{SNeIaprojected}, since the best-fit parameters minimize the merit function. Then, $\chi^{2}_{\rm min}$ gives an indication of the goodness of fit: the smaller its value, the better is the fit.

\section{Results and discussion\label{sec:results}}

The two models are contrasted with OHD and SNe Ia data through their corresponding Hubble parameters. For the barotropic model, we consider the solution~\eqref{HToyModel} subject to the Friedmann's constraint~\eqref{Friedmannconstraint}. Hence, their free parameters are $\theta=\left\{\Omega_{0m}, h, x\right\}$. For $\Omega_{0m}$ we use the flat prior $F\in[0,1]$, for $h$ we use the Gaussian prior $G(0.7324,0.0174)$, and for $x$ an EoS-like flat prior $F\in[-1,1]$. 

For the CSL model, on the other hand, we take into account all the solutions with phase transition, subject to the constraint~\eqref{FriedmannconstraintCSL10}.  Thus, the free parameters of the first two cases are $\theta=\left\{\Omega_{0m},\Omega_{0\xi}, h\right\}$, while for the last one we have $\theta=\left\{\Omega_{0m},h \right\}$. The same priors as for the barotropic model are used for $\Omega_{0m}$ and $h$. For $\Omega_{0\xi}$, we consider a flat prior $F\in[0,1]$ derived from Eq.~\eqref{FriedmannconstraintCSL1}, where the region allowed by laboratory experiments is contained. In each case, the Hubble parameter as a function of the redshift is obtained numerically.


For further comparison, we obtain the best fit of the free parameters of the $\Lambda$CDM model after the Friedmann's constraint, i.e. $\theta=\left\{\Omega_{0m},h\right\}$, with a flat prior $F\in[0,1]$, using 
\begin{align}\label{lambdaCDM}
\frac{H_{\Lambda \text{CDM}}^2}{H_0^2} &= \Omega_{0m}\left(1+z\right)^{3}+1-\Omega_{0m}.
\end{align}
It is worth mentioning that \textbf{a} parameter in the \textit{emcee} code is modified to obtain a mean acceptance fraction between $0.2$ and $0.5$ for each model~\cite{Foreman_emcee_2013}. Their values are 7 for $\Lambda$CDM, 5 for the barotropic model, and 4 for the CSL model.

Even though the value of $\chi_{\rm min}^{2}$ characterizes the goodness of the fit, it does not take into account the number of free parameters. Thus, one could add more of them at will such that the likelihood gets minimized. To overcome this problem, we compare their goodness through two statistical indicators: the Akaike Information Criterion (AIC)~\cite{Akaike_New_1974} and the Bayesian Information Criterion (BIC)~\cite{Schwarz_Estimating_1978}, defined, respectively as
\begin{align} \label{AIC}
\text{AIC}&=2\theta_{N}-2\ln{\left(\mathcal{L}_{\rm max}\right)}, \\
 \label{BIC}
\text{BIC}&=\theta_{N}\ln{\left(n\right)}-2\ln{\left(\mathcal{L}_{\rm max}\right)}.
\end{align}
Here, $\mathcal{L}_{\rm max}$ represents the maximum value of the likelihood function calculated for the best-fit parameters, $\theta_{N}$ is the number of free parameters, and $n$ is the number of the data samples. Thereby, these indicators characterize the model according to the number of free parameters, where the most favored one minimize their AIC and BIC values. Since the latter depends on the logarithm of the total observational data, it gives a better smoking gun than AIC. Thus, a model possessing a higher value of BIC when compared to the other is considered as a criterion against it.

From the observational data of OHD, SNe Ia, and their joint analysis, we provide the best-fit values for the parameters of each model in Table~\ref{bestfittable}, alongside their AIC and BIC indicators. In Figs.~\ref{TriangleLCDM}-\ref{TriangleCSLiii}, we show the allowed regions of the parameter space for the $\Lambda$CDM, barotropic, and CSL models, respectively.

The results of Table~\ref{bestfittable} indicate that, for the joint analysis, the barotropic model has lower $\chi^2_{\rm min}$ than $\Lambda$CDM, reproducing better the cosmological data of SNe Ia and OHD. However, since $\Lambda$CDM has less free parameters, it gives a lower BIC. Thus, even though we found evidence that $\Lambda$CDM is statistically preferred, the small difference between their indicators is not strong enough to be conclusive about which model is better. In this case, the age of the Universe obtained from Eq.~\eqref{agetoy} is $\tau= 14.392\pm0.183\,\si{Gyr} $.

On the other hand, the CSL model with $\Omega_{0\Lambda_{\rm eff}} > \Omega_{0\xi}$ also reproduces better the cosmological data than $\Lambda$CDM, as it can be seen from the difference of their $\chi_{\rm min}^{2}$ in the joint analysis. The $\Lambda$CDM model, in contrast, is statistically preferred as one concludes from their BIC's values. However, the best fit of $\Omega_{0\xi}$ is translated into
\begin{align}\label{valuexi}
 4.2\times10^{-19}\si{s^{-1}}\leq \xi_{\rm CSL}\leq 9.7\times10^{-19}\si{s^{-1}},   
\end{align}
which is clearly incompatible with previous laboratory experiments  (see~\cite{Josset:2016vrq} and references therein). This is in agreement with the conclusions of Ref.~\cite{Martin:2019jye} whose results obtained from the CMB are also in tension with earlier data. Moreover, imposing a prior compatible with preexisting data yields to similar values for $h$ and $\Omega_{0m}$ when compared to $\Lambda$CDM, differing only in their associated errors. This indicates that the CSL model tends to the latter when the value of $\Omega_{0\xi}\to0$. Nevertheless, the obtained value for $\Omega_{0\xi}$ is not a best fit in this case, since it does not maximize the probability within the restricted prior. The age of the Universe in this case is $\tau=14.338\pm0.183\,\si{Gyr}$. 

\begin{widetext}
\begin{center}
\begin{table}[H]
\centering
\resizebox{15cm}{!}{
\begin{tabular}{|c|cccc|ccc|}
\hline
 Data & \multicolumn{4}{c|}{Best-fit values} & \multicolumn{3}{c|}{Goodness of fit} \\
\cline{2-8}
     & $\Omega_{0m}$ & $h$ & $x$ & $\Omega_{0\xi}$ & $\chi^{2}_{\rm min}$ & AIC & BIC \\
\hline
\multicolumn{8}{|c|}{$\Lambda$CDM Model} \\
\cline{1-8}
OHD    & $0.248_{-0.014}^{+0.015}$ & $0.715_{-0.010}^{+0.010}$ & - & - & $27.9$ & $31.9$ & $35.7$ \\
SNe Ia & $0.299_{-0.021}^{+0.022}$ & $0.732_{-0.017}^{+0.017}$ & - & - & $1026.9$ & $1030.9$ & $1040.8$ \\
Joint  & $0.265_{-0.012}^{+0.013}$ & $0.705_{-0.009}^{+0.009}$ & - & - & $1057.1$ & $1061.1$ & $1071.1$ \\
\hline
\multicolumn{8}{|c|}{Barotropic Model} \\
\hline
OHD    & $0.267_{-0.035}^{+0.038}$ & $0.710_{-0.014}^{+0.014}$ & $0.027_{-0.045}^{+0.046}$ & - & $27.0$ & $33.0$ &  $38.8$ \\
SNe Ia & $0.306_{-0.050}^{+0.058}$ & $0.732_{-0.017}^{+0.017}$ & $0.041_{-0.194}^{+0.310}$ & - & $1027.1$ & $1033.1$ & $1048.0$ \\
Joint  & $0.295_{-0.022}^{+0.023}$ & $0.699_{-0.009}^{+0.009}$ & $0.054_{-0.032}^{+0.035}$ & - & $1053.5$ & $1059.5$ & $1074.5$ \\  
\hline
\multicolumn{8}{|c|}{CSL Model with $\Omega_{0\Lambda_{\rm eff}}>\Omega_{0\xi}$}\\
\hline
OHD    & $0.301_{-0.031}^{+0.038}$ & $0.700_{-0.013}^{+0.012}$ & - & $0.009_{-0.006}^{+0.013}$ & $26.7$ &                     $32.7$ & $38.5$ \\
SNe Ia & $0.346_{-0.032}^{+0.031}$ & $0.732_{-0.017}^{+0.018}$ & - & $0.052_{-0.038}^{+0.056}$ & $1028.2$ &                   $1034.2$ & $1049.1$ \\
Joint  & $0.310_{-0.022}^{+0.025}$ & $0.697_{-0.009}^{+0.009}$ & - & $0.010_{-0.006}^{+0.010}$ & $1053.7$ &                   $1059.7$ & $1074.7$ \\
\hline
\multicolumn{8}{|c|}{CSL Model with $\Omega_{0\Lambda_{\rm eff}}<\Omega_{0\xi}$}\\
\hline
OHD    & $0.487_{-0.020}^{+0.023}$ & $0.659_{-0.009}^{+0.009}$ & - & $0.106_{-0.010}^{+0.013}$ & $31.7$ & $37.7$ & $43.5$ \\
SNe Ia & $0.415_{-0.031}^{+0.041}$ & $0.732_{-0.017}^{+0.017}$ & - & $0.237_{-0.069}^{+0.157}$ & $1031.5$ & $1037.5$ & $1052.4$ \\
Joint  & $0.450_{-0.014}^{+0.015}$ & $0.674_{-0.008}^{+0.008}$ & - & $0.116_{-0.008}^{+0.009}$ & $1078.4$ & $1084.4$ & $1099.4$ \\
\hline
\multicolumn{8}{|c|}{CSL Model with $\Omega_{0\Lambda_{\rm eff}}=\Omega_{0\xi}$}\\
\hline
OHD    & $0.474_{-0.016}^{+0.016}$ & $0.662_{-0.009}^{+0.009}$ & - & - & $31.1$ & $35.1$ & $39.0$ \\
SNe Ia & $0.385_{-0.022}^{+0.023}$ & $0.732_{-0.017}^{+0.017}$ & - & - & $1029.9$ & $1033.9$ & $1043.8$ \\
Joint  & $0.446_{-0.013}^{+0.013}$ & $0.675_{-0.008}^{+0.008}$ & - & - & $1076.8$ & $1080.8$ & $1090.8$ \\
\hline
\end{tabular}}
\caption{Results of best-fit parameters and statistical indicators. The uncertainties correspond to $1\sigma$ ($68.3\%$) of confidence level (CL). } \label{bestfittable}
\end{table}   
\end{center}
\end{widetext}

The cases when $\Omega_{0\Lambda_{\rm eff}}<\Omega_{0\xi}$ and $\Omega_{0\Lambda_{\rm eff}} = \Omega_{0\xi}$ do not minimize neither the $\chi^2_{\rm min}$ nor the BIC's value in comparison with $\Lambda$CDM. Indeed, although the latter case has less free parameters than the barotropic and CSL model with $\Omega_{0\Lambda_{\rm eff}}>\Omega_{0\xi}$, the large value of its $\chi^2_{\rm min}$ undermines the remaining statistical indicators. Even more, in both cases, the best-fit values of $\Omega_{0m}$ obtained from the joint analysis of OHD and SNe Ia data are ruled out by model-independent estimations~\cite{Holanda_2019}. Thus, we conclude that these two cases are not well supported by cosmological observations.

\section{Conclusions and further remarks\label{sec:conclusions}}

In this work, we address the issue of higher-order equations generated by energy diffusion in homogeneous and isotropic UG~\cite{Garcia-Aspeitia:2019yni,Garcia-Aspeitia:2019yod}, by assuming an EoS that relates the diffusion function with the energy density of matter. It is worth recalling that diffusion is compatible with UG by virtue of its restricted diffeomorphism invariance, induced by the presence of the nondynamical $4$-form $\varepsilon_0(x)$, whose value is constrained through the cosmic history through Eq.~\eqref{varepsflrw}. We study a Universe filled with dark matter that develops energy diffusion and, to determine their observational viability, two models possessing such a feature are considered and contrasted with cosmological data. 

First, we study a barotropic model that incorporates the phenomenon of diffusion to gain intuition. Although the cosmological integration constant is fixed by initial data, energy diffusion introduces an additional parameter. Four exact solutions are found in this scenario, with one of them having a phase transition from decelerated to accelerated expansion. Remarkably, this model is successful at describing the joint analysis of OHD and SNe Ia data, even though $\Lambda$CDM is statistically preferred for having less free parameters. However, their small difference in BIC's value indicates that, in order to be conclusive, more data are needed.  

Second, a well-known scenario with energy-momentum nonconservation generated by quantum collapse is analyzed: the CSL model. Three analytical homogeneous and isotropic solutions are found. Remarkably, all of them possess phase transition induced by energy diffusion. When contrasted with observations, the CSL model with $\Omega_{0\Lambda_{\rm eff}}>\Omega_{0\xi}$ gives, according to experimental data, a better description of the cosmological evolution than $\Lambda$CDM model. Nevertheless, the latter has fewer free parameters and it is therefore more favorable from a statistical viewpoint. However, the best-fit value of $\Omega_{0\xi}$ forces the free parameter of the CSL model to lie in a region that is incompatible with other laboratory experiments. On the other hand, the cases when $\Omega_{0\Lambda_{\rm eff}}=\Omega_{0\xi}$ and $\Omega_{0\Lambda_{\rm eff}}<\Omega_{0\xi}$ give neither a good fit nor a compatible prediction with model-independent measurements of $\Omega_{0m}$.

Even though we found evidence that some cases studied here tend to $\Lambda$CDM asymptotically, the source of the accelerated expansion is radically different in scenarios with diffusion. Interestingly enough, the CSL model that emulates a variable cosmological constant with sign flip at late times opens the possibility of having a Universe with phase transition from de Sitter to anti de Sitter expansion with a very far future cosmological singularity: a big crunch. The absence of turning points in the Hubble parameter, as predicted by $\Lambda$CDM and supported by experiments, is a remarkable feature of the aforementioned models possessing a sign flip in the variable cosmological constant. This may be an attractive consequence regarding the swampland criteria on cosmology~\cite{Agrawal:2018own}. Although different scenarios with variable cosmological constant have been proposed beyond $\Lambda$CDM~\cite{Caldwell:1997ii,Clifton:2011jh,vanPutten:2015wma,Colgain:2018wgk,10.1093/mnrasl/slz158}, diffusion provides a rich phenomenology dispensing from additional fields and higher-curvature corrections to the Einstein--Hilbert action.

It is worth mentioning that the constraints obtained here are model dependent. Indeed, we found that diffusion is not ruled out by SNe Ia or Hubble data in the barotropic model, providing an appealing approach to describe late-time acceleration. This may open new avenues to be explored, for instance, in the context of cosmological perturbations. In Ref.~\cite{Gao:2014nia}, it was shown that UG could have observable effects in the CMB through the Sachs--Wolfe effect~\cite{Sachs:1967er}. For adiabatic fluctuations, however, the differences with GR are negligible at large scales. Later, the same effect was computed in gauge-invariant way, concluding that cosmological perturbations of UG and GR are identical~\cite{Basak:2015swx}. Nevertheless, both studies assumed the conservation of the energy-momentum tensor and it is certainly of great interest to study the effect of energy diffusion to determine whether it could affect the CMB or structure formation. 

On the other hand, the only viable case of the CSL model, namely $\Omega_{0\Lambda_{\rm eff}}>\Omega_{0\xi}$, is in tension with previous constraints, since the best-fit values of $\xi_{\rm CSL}$ in Eq.~\eqref{valuexi} are incompatible with~\cite{TOROS20173921} (see also~\cite{Josset:2016vrq}). The nature of this incompatibility is different from Ref.~\cite{Martin:2019jye}, whose limitations in the inflationary epoch were recently pointed out in~\cite{Bengochea:2020qsd}. The extrapolation in this work is far less dramatic than~\cite{Martin:2019jye}, since the effects of diffusion start from hadronization to the present time. Of course, there is still the possibility that the effective parameters changed after nucleosynthesis, affecting the estimations done here. This fact reinforces the idea of keep exploring the implications of energy diffusion in cosmology, in particular collapse models.


An appealing approach to restore the energy-momentum conservation in UG is to consider interacting fluids. Although their energy-momentum tensor may not need to be conserved individually, their interaction could imply a energy conservation of the whole system.\footnote{We thank A. Arza for pointing this out.} This setup has been considered in different scenarios~\cite{Chimento:2005wv,Cataldo:2008tx,Farajollahi:2010rd,Lip:2010dr,Cotsakis:2012sh,Elizalde:2017dmu,Cruz:2019kgz}, but it has not been implemented in UG. The cosmological consequences of such interaction are worth exploring, since the backreaction of more than one fluid becomes relevant in phase transitions between different matter dominated eras. The predictions of the interacting model can be contrasted with observational data in the same spirit as it has been done here. These studies are relevant for determining the viability of UG with energy-momentum nonconservation and they are left for future contributions.

\begin{acknowledgments}
The authors thank to Ariel Arza, Yuri Bonder, Eoin \'O. Colg\'ain, Daniel Sudarsky, Maurice H.P.M. van Putten, and Nelson Videla for valuable comments. C.C. thanks the hospitality of Universidad de Santiago de Chile as well as the partial support of Proyecto POSTDOC\_DICYT, C\'odigo 041931CM\_POSTDOC, Universidad de Santiago de Chile. N.C. is partially supported by the Direcci\'on de Investigaci\'on Cient\'ifica y Tecnol\'ogica (Dicyt) from Universidad de Santiago de Chile through Grant N$^\circ$ 041831CM. 
\end{acknowledgments}

\begin{widetext}

\begin{figure}[H]
\centering
\includegraphics[scale=0.4]{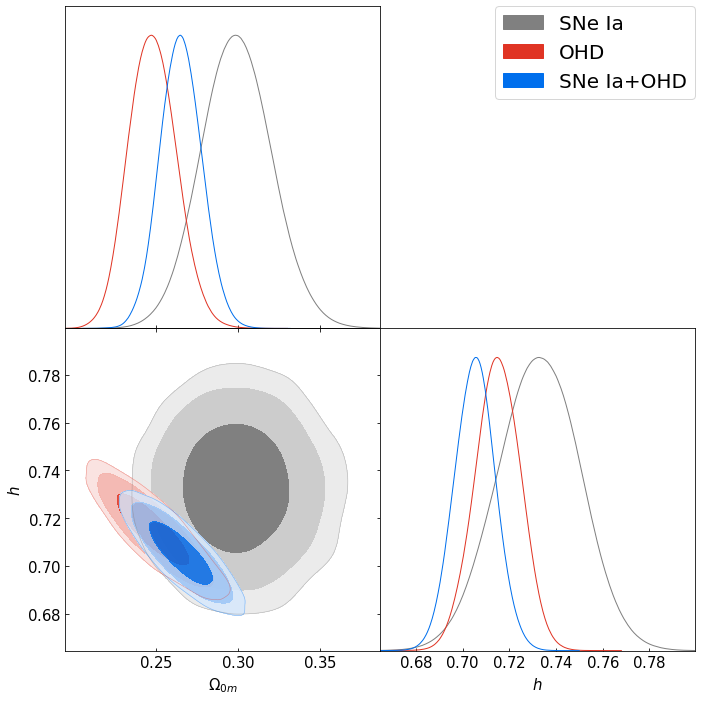}
\caption{Joint and marginalized constraints of $\Omega_{0m}$ and $h$ for the $\Lambda$CDM model,  using the Bayesian statistical analysis of Sec.~\ref{sec:constraints}. The admissible regions correspond to $1\sigma\left(68.3\%\right)$, $2\sigma\left(95.5\%\right)$, and $3\sigma\left(99.7\%\right)$ CL, respectively. The best-fit values in this case, obtained from the joint analysis of SNe Ia and OHD, are $\Omega_{0m}=0.265
^{+0.013}_{-0.012}$ and $h=0.705^{+0.009}_{-0.009}$. These values are used for the sake of comparison with other models.}
\label{TriangleLCDM}
\end{figure}

\begin{figure}[H]
\centering
\includegraphics[scale=0.4]{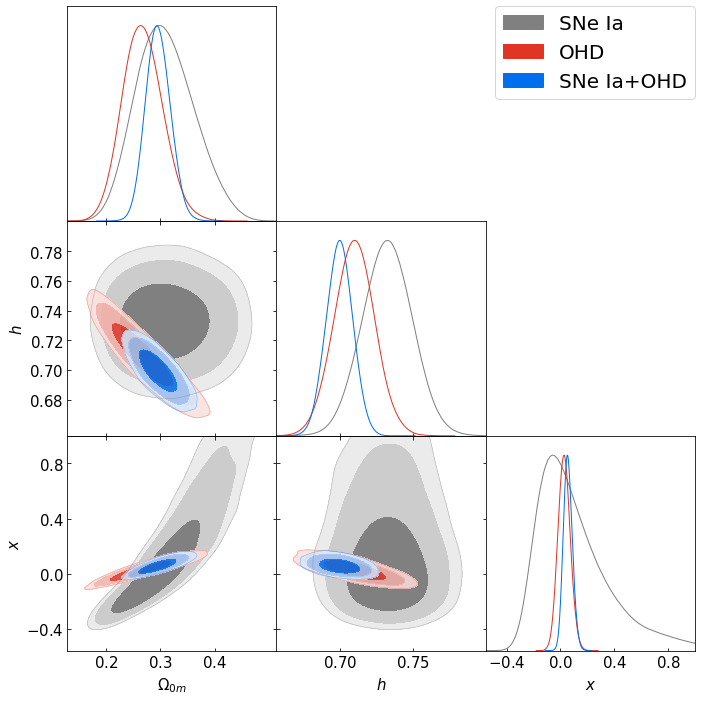}
\caption{Joint and marginalized constraint of $\Omega_{0m}$, $h$ and $x$ for the barotropic model, using the Bayesian statistical analysis of Sec.~\ref{sec:constraints}. Recall, this model assumes a barotropic EoS relating the diffusion function with energy density through $Q=x\rho$, where $x$ is a dimensionless free parameter. The admissible regions correspond to $1\sigma\left(68.3\%\right)$, $2\sigma\left(95.5\%\right)$, and $3\sigma\left(99.7\%\right)$ CL, respectively. The best-fit values for this model, derived from the joint analysis of SNe Ia and OHD, are $\Omega_{0m}=0.295^{+0.023}_{-0.022}$, $h=0.699^{+0.009}_{-0.009}$, and $x=0.054^{+0.035}_{-0.032}$. In principle, this model describes the late-time acceleration at the background level and more analysis is needed to test its viability in further regimes.}
\label{TriangleToy}
\end{figure}

 \begin{figure}[H]
 \centering
 \includegraphics[scale=0.4]{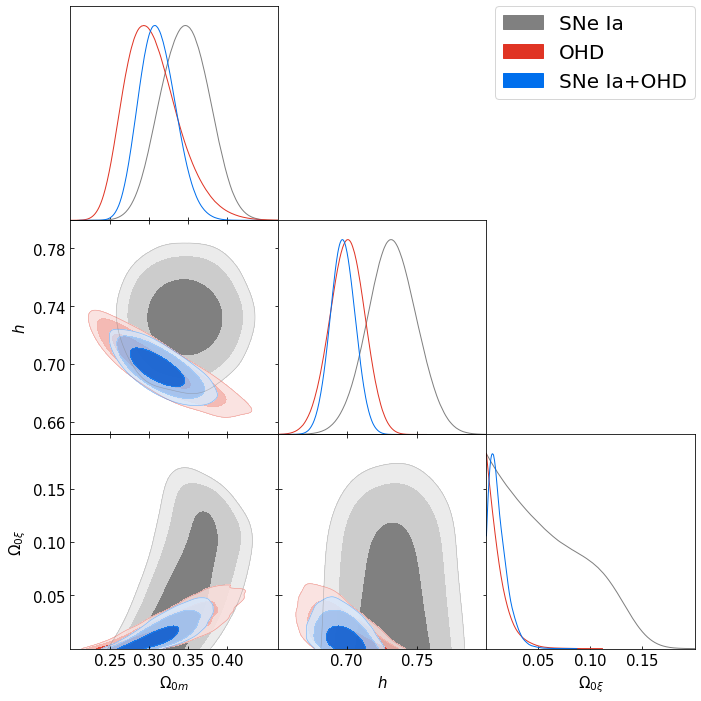}
 \caption{Joint and marginalized constraint of $\Omega_{0m}$, $h$ and $\Omega_{0\xi}$ for the CSL model with $\Omega_{0\Lambda_{\rm eff}}>\Omega_{0\xi}$, using the Bayesian statistical analysis of Sec.~\ref{sec:constraints}. The relation between the diffusion function and energy density can be expressed as $\dot{Q}=-\xi_{\rm CSL}\rho$, where dot denotes derivative with respect to cosmic time. The admissible regions correspond to $1\sigma\left(68.3\%\right)$, $2\sigma\left(95.5\%\right)$, and $3\sigma\left(99.7\%\right)$ CL, respectively.  The joint analysis of SNe Ia in this case provides best-fit values parameters $\Omega_{0m}=0.310^{+0.025}_{-0.022}$, $h=0.697^{+0.009}_{-0.009}$, and $\Omega_{0\xi}=0.010^{+0.010}_{-0.006}$. The latter can be translated into constraints on $\xi_{\rm CSL}$ given in Eq.~\eqref{valuexi}, which are in tension with~\cite{TOROS20173921} (see also~\cite{Josset:2016vrq}).}
 \label{TriangleCSL}
 \end{figure}
 
 \begin{figure}[H]
 \centering
 \includegraphics[scale=0.4]{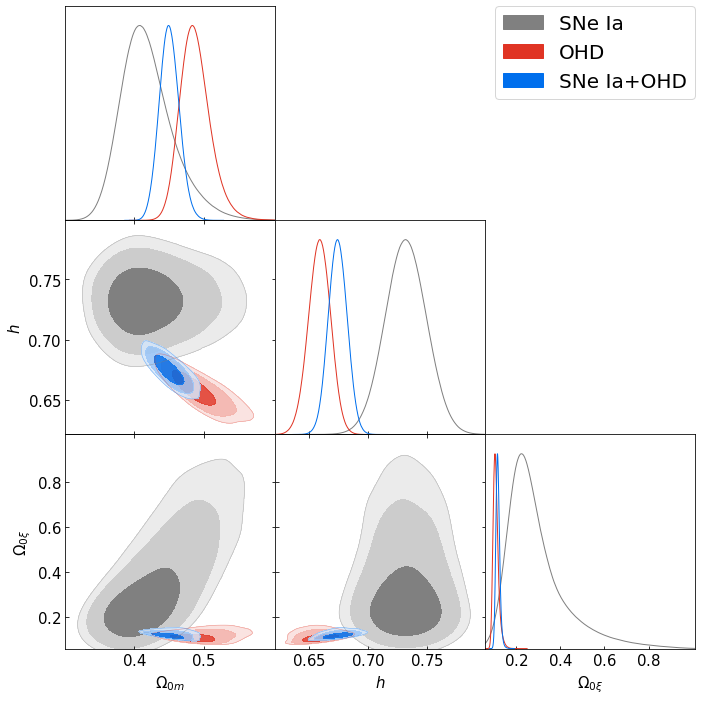}
 \caption{Joint and marginalized constraint of $\Omega_{0m}$, $h$ and $\Omega_{0\xi}$ for the CSL model with $\Omega_{\Lambda_{0\rm eff}}<\Omega_{0\xi}$, using the Bayesian statistical analysis of Sec.~\ref{sec:constraints}. Recall, the relation of the diffusion function and energy density for the CSL model is given in Eq.~\eqref{diffusioncsl}. Moreover, this case predicts a sign flip of the variable cosmological constant~\eqref{variablelambda} and big crunch at very far future. The admissible regions correspond to $1\sigma\left(68.3\%\right)$, $2\sigma\left(95.5\%\right)$, and $3\sigma\left(99.7\%\right)$ CL, respectively. The best-fit values obtained from the joint analysis of SNe Ia and OHD are $\Omega_{0m}=0.450^{+0.015}_{-0.014}$, $h=0.674^{+0.008}_{-0.008}$, and $\Omega_{0\xi}=0.116^{+0.009}_{-0.008}$. However, the indicators $\chi_{\rm min}^2$, $\mbox{AIC}$, and $\mbox{BIC}$, are not minimized with respect to $\Lambda$CDM, showing that this model is not statistically preferred.}
 \label{TriangleCSLii}
 \end{figure}

\begin{figure}[H]
\centering
\includegraphics[scale=0.36]{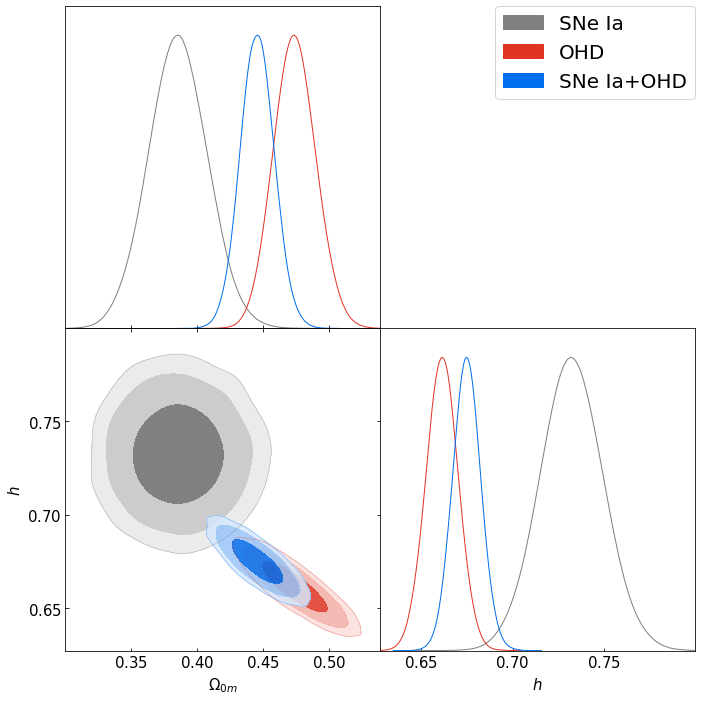}
\caption{Joint and marginalized constraint of $\Omega_{0m}$ and $h$ for the CSL model with $\Omega_{0\Lambda_{\rm eff}}=\Omega_{0\xi}$, using the Bayesian statistical analysis of Sec.~\ref{sec:constraints} and the relation between diffusion and energy density of Eq.~\eqref{diffusioncsl}. This case has one less free parameter than the other two possibilities in the CSL model by virtue of $\Omega_{0\Lambda_{\rm eff}}=\Omega_{0\xi}$. The admissible regions correspond to $1\sigma\left(68.3\%\right)$, $2\sigma\left(95.5\%\right)$, and $3\sigma\left(99.7\%\right)$ CL, respectively. Joint analysis of SNe Ia and OHD data gives best-fit value parameters $\Omega_{0m} = 0.446^{+0.013}_{-0.013}$ and $h=0.675^{+0.008}_{-0.008}$. However, this case is ruled out by model-independent estimations of matter density~\cite{Holanda_2019}.}
\label{TriangleCSLiii}
\end{figure}

\end{widetext}

\bibliography{References}

\end{document}